\documentclass[twocolumn,twoside,slac_two]{revtex4}
\usepackage{graphicx}
\usepackage{fancyhdr}
\def\be{\begin{equation}}
\def\ee{\end{equation}}
\def\bc{\begin{center}}
\def\ec{\end{center}}
\def\bea{\begin{eqnarray}}
\def\eea{\end{eqnarray}}
\def\dd{\displaystyle}
\def\nn{\nonumber}

\def\ov{\overline}
\def\gappeq{\mathrel{\rlap {\raise.5ex\hbox{$>$}} {\lower.5ex\hbox{$\sim$}}}}
\def\lappeq{\mathrel{\rlap{\raise.5ex\hbox{$<$}} {\lower.5ex\hbox{$\sim$}}}}
\def\GeV{{\rm GeV}}

\begin{document}

\hfill{~~~~~~~~~~~RM3-TH/06-18}

\hfill{~~~~~~~~~~~CERN-PH-TH/2006-206}

\vskip 1cm

\title{An Update on Models of Neutrino Masses and Mixings }

\author{Guido Altarelli}
\affiliation{Dipartimento di Fisica `E.~Amaldi', Universit\`a di Roma Tre
and INFN, Sezione di Roma Tre, I-00146 Rome, Italy and \\ CERN, Department of Physics, Theory Division, 
 CH--1211 Geneva 23, Switzerland}

\begin{abstract}
At the School I gave three lectures on neutrino masses and mixings. Much of the material covered in my first two lectures is written down in a review on the subject that I published not long ago with F. Feruglio \cite{rev}. Here, I make a relatively short summary (with updates) of the content of my first two lectures, referring to our review for a more detailed presentation, and then I expand on the content of the third lecture which was dedicated to recent work on A4 models of tri-bimaximal neutrino mixing which were not covered in the review.
\end{abstract}

\maketitle

\thispagestyle{fancy}

\section{Introduction} 
By now there is convincing evidence for solar and atmospheric neutrino oscillations. The $\Delta m^2$ values and mixing angles are known with fair accuracy.  A summary, taken from Ref. \cite{one} of the results is shown in Table~\ref{tab01}. For the $\Delta m^2$ we have:  $\Delta m^2_{atm}\sim 2.4~10^{-3}~eV^2$ and  $\Delta m^2_{sol}\sim ~7.9~10^{-5}~eV^2$. As for the mixing angles, two are large and one is small. The atmospheric angle $\theta_{23}$ is large, actually compatible with maximal but not necessarily so: at $3\sigma$: $0.29 \lappeq \sin^2{\theta_{23}}\lappeq 0.71$ with central value around  $0.44$. The solar angle $\theta_{12}$, the most precisely measured, is large, $\sin^2{\theta_{12}}\sim 0.31$, but certainly not maximal (by about 6 $\sigma$ now). The third angle $\theta_{13}$, strongly limited mainly by the CHOOZ experiment, has at present a $3\sigma$ upper limit given by about $\sin^2{\theta_{13}}\lappeq 0.04$.

\begin{table*}[htb]
\begin{center}
\caption[]{Best fit values of squared mass differences and mixing angles\cite{one}}
\label{tab01}
\begin{tabular}{|c|c|c|c|}   
\hline  
& & & \\   
&{\tt lower limit} & {\tt best value} & {\tt upper limit}\\
&($2\sigma$)& & ($2\sigma$)\\
\hline
$(\Delta m^2_{sun})_{\rm LA}~(10^{-5}~{\rm eV}^2)$ & 7.2& 7.9&8.6\\
\hline
$\Delta m^2_{atm}~(10^{-3}~{\rm eV}^2)$ & 1.8& 2.4& 2.9\\
\hline
$\sin^2\theta_{12}$ & 0.27 & 0.31 &0.37\\
\hline
$\sin^2\theta_{23}$ & 0.34 & 0.44 &0.62\\
\hline
$\sin^2\theta_{13}$ & 0 & 0.009 &0.032\\
\hline
\end{tabular} 
\end{center}
\end{table*}

In spite of this experimental progress there are still many alternative routes in constructing models of neutrino masses. This variety is mostly due to the
considerable ambiguities that remain. First of all, it is essential to know whether
the LSND signal, which has not been confirmed by KARMEN and is currently being double-checked by MiniBoone, will be confirmed or will be excluded.  If LSND is right we probably
need at least four light neutrinos; if not we can do with only the three known ones, as we assume here in the following.  Then, as
neutrino oscillations only determine mass squared differences a crucial missing input is the absolute scale of neutrino masses. Also the pattern of the neutrino mass spectrum is not known: it could be approximately degenerate with $m^2>> \Delta m^2_{ij}$ or of the inverse hierarchy type (with the 1,2 solar doublet on top) or of the normal hierarchy type (with the solar doublet below).

The following experimental information on the absolute scale of neutrino masses is available. From the endpoint of tritium beta decay spectrum we have  an absolute upper limit of
2 eV (at 95\% C.L.) on the mass of "$\bar \nu_e$" \cite{PDG}, which, combined with the observed oscillation
frequencies under the assumption of three CPT-invariant light neutrinos, represents also an upper bound on the masses of
all active neutrinos. A less direct information on the mass scale is obtained from neutrinoless double beta decay ($0\nu \beta \beta$). The discovery of $0\nu \beta \beta$ decay would be very important because it would directly establish lepton number violation and
the Majorana nature of $\nu$'s. 
The present limit from $0\nu \beta \beta$ is affected by a relatively large uncertainty due to ambiguities on nuclear matrix elements. We quote here two recent limits ($90\%$c.l.): $\vert m_{ee}\vert< 0.33-1.35$ eV  [IGEX($^{76}Ge$) \cite{igex}] or 
$\vert m_{ee}\vert < (0.2-1.1)$ eV [Cuoricino($^{130}Te$) \cite{cuo}], where $ m_{ee} =  \sum{U_{ei}^2 m_i}$ in terms of the mixing matrix and the mass eigenvalues (see eq.(\ref{3nu1gen})). Complementary information on the sum of neutrino masses is also provided by measurements in cosmology, where an extraordinary progress has been made in the last years, in particular data on the cosmic  microwave background (CMB)anisotropies (WMAP), on the large scale structure of the mass distribution in the Universe (SDSS, 2dFGRS) and from the  Lyman alpha forest \cite{lepa}. WMAP by itself is not very restrictive: 
$\sum_i \vert m_i\vert < 2.11$ eV (at 95\% C.L.).  Combining CMB data  with those on the large scale structure one obtains $\sum_i \vert m_i\vert < 0.68$ eV. Adding also the data from the Lyman alpha forest one has $\sum_i \vert m_i\vert < 0.17$ eV \cite{sel}. But this last combination is questionable because of some tension (at $\sim 2\sigma$'s) between the Lyman alpha forest data and those on the large scale structure. In any case, the cosmological bounds depend on a number of assumptions (or, in fashionable terms, priors) on the cosmological model. In summary, from cosmology for 3 degenerate neutrinos of mass $m$, depending on which data sets we include and on our degree of confidence in cosmological models, we can conclude that $m \lappeq
0.06-0.23-0.7~eV$.

Given that neutrino masses are certainly extremely
small, it is really difficult from the theory point of view to avoid the conclusion that L conservation is probably  violated.
In fact, in terms of lepton number violation the smallness of neutrino masses can be naturally explained as inversely proportional
to the very large scale where L is violated, of order $M_{GUT}$ or even $M_{Pl}$.
If neutrinos are Majorana particles, their masses arise  from the generic dimension-five non renormalizable operator of the form: 
\be 
O_5=\frac{(H l)^T_i \lambda_{ij} (H l)_j}{M}+~h.c.~~~,
\label{O5}
\ee  
with $H$ being the ordinary Higgs doublet, $l_i$ the SU(2) lepton doublets, $\lambda$ a matrix in  flavour space,
$M$ a large scale of mass and a charge conjugation matrix $C$
between the lepton fields is understood. 
Neutrino masses generated by $O_5$ are of the order
$m_{\nu}\approx v^2/M$ for $\lambda_{ij}\approx {\rm O}(1)$, where $v\sim {\rm O}(100~\GeV)$ is the vacuum
expectation value of the ordinary Higgs. A particular realization leading to comparable masses is the see-saw mechanism, where $M$ derives from the exchange of heavy $\nu_R$'s: the resulting neutrino mass matrix reads:
\be  m_{\nu}=m_D^T M^{-1}m_D~~~.
\ee 
that is, the light neutrino masses are quadratic in the Dirac
masses and inversely proportional to the large Majorana mass.  For
$m_{\nu}\approx \sqrt{\Delta m^2_{atm}}\approx 0.05$ eV and 
$m_{\nu}\approx m_D^2/M$ with $m_D\approx v
\approx 200~GeV$ we find $M\approx 10^{15}~GeV$ which indeed is an impressive indication for
$M_{GUT}$. Thus probably neutrino masses are a probe into the physics at $M_{GUT}$. 

\section{The $\nu$-Mixing Matrix}

If we take maximal $s_{23}$  ($s_{ij}= \sin  \theta_{ij} $) and keep only linear terms in $u=  s_{13}e^{i\varphi}$, from experiment we find the following
structure of the
$U_{fi}$ ($f=e$,$\mu$,$\tau$, $i=1,2,3$) mixing matrix, apart from sign convention redefinitions: 
\bea  \label{ufi1}
&&U_{fi}= \nonumber\\
&&\left(\matrix{ c_{12}&s_{12}&u \cr  -(s_{12}+c_{12}u^*)/\sqrt{2}&(c_{12}-s_{12}u^*)/\sqrt{2}&1/\sqrt{2}\cr
(s_{12}-c_{12}u^*)/\sqrt{2}&-(c_{12}+s_{12}u^*)/\sqrt{2}&1/\sqrt{2}     } 
\right)\nonumber\\
\eea
If $s_{13}$ would be exactly zero there would be no CP violations in $\nu$ oscillations. A main target of the new planned oscillation experiments is to measure the actual size of $s_{13}$. In the next decade the upper limit on $sin^2{2\theta_{13}}$ will possibly go down by at least an order of magnitude  (T2K, No$\nu$A, DoubleCHOOZ.....) \cite{lecsmi}. Even for three neutrinos the pattern of the neutrino mass spectrum is still undetermined: it can be approximately degenerate, or of the inverse hierarchy type or normally hierarchical. Given the observed frequencies and  the notation $\Delta m^2_{sun}\equiv \Delta m^2_{12}$,
$\Delta m^2_{atm}\equiv \vert\Delta m^2_{23}\vert$ with
$\Delta m^2_{12}=\vert m_2\vert^2-\vert m_1\vert^2 > 0$ and $\Delta m^2_{23}= m_3^2-\vert m_2\vert ^2$, the three possible patterns of mass eigenvalues are:
\bea \label{abc}
&&{\tt{Degenerate}}:  |m_1|\sim |m_2| \sim |m_3|\gg |m_i-m_j|\nonumber\\ 
&&{\tt{Inverted~hierarchy}}:  |m_1|\sim
|m_2| \gg |m_3| \nonumber\\ 
&&{\tt{Normal~hierarchy} }:  |m_3| \gg |m_{2,1}|
\eea  
The sign of $\Delta m^2_{23}$ can be measured in the future through matter effects in long baseline experiments \cite{lecsmi}. Models based on all these patterns have been proposed and studied and all are in fact viable at present.

The detection of neutrino-less double beta decay, besides its enormous intrinsic importance as a direct evidence of $L$ non conservation, would also offer a way to possibly disintangle the 3 cases.  The quantity which is bound by experiments
is the 11 entry of the
$\nu$ mass matrix, which in general, from $m_{\nu}=U^* m_{diag} U^\dagger$, is given by :
\bea 
\vert m_{ee}\vert~=\vert(1-s^2_{13})~(m_1 c^2_{12}~+~m_2 s^2_{12})+m_3 e^{2 i\phi} s^2_{13}\vert
\label{3nu1gen}
\eea
Starting from this general formula it is simple to
derive the following bounds for degenerate, inverse hierarchy or normal hierarchy mass patterns.
\begin{itemize}
\item[a)]  Degenerate case. If $|m|$ is the common mass and we set $s_{13}=0$, which is a safe
approximation in this case, because $|m_3|$ cannot compensate for the smallness of $s_{13}$, we have
$m_{ee}\sim |m|(c_{12}^2\pm s_{12}^2)$.  Here the phase ambiguity has been reduced to a sign ambiguity which is sufficient
for deriving bounds.  So, depending on the sign we have
$m_{ee}=|m|$ or
$m_{ee}=|m|cos2\theta_{12}$. We conclude that in  this case $m_{ee}$ could be as large as the present experimental limit
but should be at least of
order $O(\sqrt{\Delta m^2_{atm}})~\sim~O(10^{-2}~ {\rm eV})$ given that the solar angle cannot be too close to maximal (in which case
the minus sign option could be arbitrarily small). The experimental 2-$\sigma$ range of the solar angle does not
favour a cancellation by more than a factor of about 3.
\item[b)]  Inverse hierarchy case. In this case the same approximate formula $m_{ee}=|m|(c_{12}^2\pm s_{12}^2)$ holds 
because $m_3$ is small and the $s_{13}$ term in eq.(\ref{3nu1gen}) can be neglected. The difference is that here we know that $|m|\approx 
\sqrt{\Delta m^2_{atm}}$ so that $\vert m_{ee}\vert<\sqrt{\Delta m^2_{atm}}~\sim~0.05$ eV. At the same time,
since a full cancellation between the two contributions cannot take place, we expect 
$\vert m_{ee}\vert > 0.01$ eV.

\item[c)]  Normal hierarchy case. Here we cannot in general neglect the $m_3$ term. However in this case $\vert
m_{ee}\vert~\sim~
\vert\sqrt{\Delta m^2_{sun}}~ s_{12}^2~\pm~\sqrt{\Delta m^2_{atm}}~ s_{13}^2\vert$ and we have the bound 
$\vert m_{ee}\vert <$ a few $10^{-3}$ eV.
\end{itemize}
Recently some evidence for $0\nu \beta \beta$ was claimed \cite{kla} corresponding to
$\vert m_{ee}\vert\sim (0.2\div 0.6)~ {\rm eV}$ ($(0.1\div 0.9)~ {\rm eV}$ in a more conservative estimate of the involved nuclear matrix elements). This result is not supported by the IGEX and Cuoricino measurements of a comparable sensitivity, but If confirmed it would rule out cases b) and c) and point to case a) or to models
with more than 3 neutrinos. In the next few years a new generation of experiments will reach a larger sensitivity on $0\nu \beta \beta$ by about an order of magnitude. If these experiments will observe a signal this would indicate that the inverse hierarchy is realized, if not, then the normal hierarchy case remains a possibility. 

\section{ "Normal" versus "Exceptional" Models}

After KamLAND, SNO and WMAP not too much hierarchy in neutrino masses is indicated by experiments: 
\bea
r = \Delta m_{sol}^2/\Delta m_{atm}^2 \sim 1/30.\label{r}
\eea
 Precisely at $2\sigma$: $0.025 \lappeq r \lappeq 0.049$ \cite{one}. Thus, for a hierarchical spectrum, $m_2/m_3 \sim \sqrt{r} \sim 0.2$, which is comparable to the Cabibbo angle $\lambda_C \sim 0.22$ or $\sqrt{m_{\mu}/m_{\tau}} \sim 0.24$. This suggests that the same hierarchy parameter (raised to powers with o(1) exponents) applies for quark, charged lepton and neutrino mass matrices. This in turn indicates that, in absence of some special dynamical reason, we do not expect a quantity like $\theta_{13}$ to be too small. Indeed it would be very important to know how small the mixing angle $\theta_{13}$  is and how close to maximal is $\theta_{23}$. Actually one can make a  distinction between "normal" and "exceptional" models. For normal models  $\theta_{23}$ is not too close to maximal and $\theta_{13}$ is not too small, typically a small power of the self-suggesting order parameter $\sqrt{r}$, with $r=\Delta m_{sol}^2/\Delta m_{atm}^2 \sim 1/30$. Exceptional models are those where some symmetry or dynamical feature assures in a natural way the near vanishing of $\theta_{13}$ and/or of $\theta_{23}- \pi/4$. Normal models are conceptually more economical and much simpler to construct. 
Typical categories of normal models are:
\begin{itemize}
\item[a)] Anarchy. These are models with approximately degenerate mass spectrum and no ordering principle, no approximate symmetry assumed in the neutrino mass sector \cite{anarchy} \cite{rev}. The small value of r is accidental, due to random fluctuations of matrix elements in the Dirac and Majorana neutrino mass matrices. Starting from a random input for each matrix element, the see-saw formula, being a product of 3 matrices, generates a broad distribution of r values. All mixing angles are generically large: so in this case one does not expect $\theta_{23}$ to be maximal and $\theta_{13}$ must probably be found near its upper bound.
\item[b)] Semianarchy. We have seen that anarchy is the absence of structure in the neutrino sector. Here we consider an attenuation of anarchy where
the absence of structure is limited to the 23 sector. The typical structure is in this case \cite{lopsu1}  \cite{rev}: 
\be m_\nu
\approx m
\left(
\begin{array}{ccc}
\delta& \epsilon&\epsilon \\
\epsilon& 1&1\\ \epsilon& 1& 1
\end{array}
\right)
\label{s-an}~~~,
\ee
where $\delta$ and $\epsilon$ are small and by 1 we mean entries of $o(1)$ and also the 23 determinant is of $o(1)$. 
This texture can be realized, for example, without see-saw from a suitable set of $U(1)_F$ charges for $(l_1, l_2,l_3)$, eg $(a,0,0)$ appearing in the dim. 5 operator of eq.(\ref{O5}). Clearly, in general we would
expect two mass eigenvalues of order 1, in units of $m$, and one small, of order $\delta$ or $\epsilon^2$. 
This typical pattern would not fit the observed solar
and atmospheric observed frequencies. However, given that $\sqrt{r}$ is not too small, we can assume that its small value is
generated accidentally, as for anarchy. We see that, if by chance the second eigenvalue $\eta\sim \sqrt{r}\sim \delta+\epsilon^2$, we can then obtain the correct value of $r$ together
with large but in general non maximal $\theta_{23}$ and $\theta_{12}$ and small $\theta_{13}\sim \epsilon$. The guaranteed smallness of $\theta_{13}$ is the main advantage over anarchy, and the relation with $\sqrt{r}$ normally keeps $\theta_{13}$ not too small. For example,  $\delta\sim \epsilon^2$ in typical $U(1)_F$ models that provide a very economical but effective realization of this scheme . 
\item[c)] Inverse hierarchy. In the limit of exact $L_e-L_{\mu}-L_{\tau}$ symmetry we have inverted hierarchy with $r=0$ and bi-maximal mixing (both  $\theta_{12}$ and  $\theta_{23}$ are maximal) \cite{rev}. Simple forms of symmetry breaking cannot sufficiently displace $\theta_{12}$ from the maximal value because typically
$\tan^2{\theta_{12}} \sim 1+o(r)$. Viable normal models can be obtained by arranging large contributions to $\theta_{23}$ and $\theta_{12}$  from the charged lepton mass diagonalization. But then, it turns out that, in order to obtain the measured value of $\theta_{12}$,  the size of $\theta_{13}$ must be close to its present upper bound \cite{chlep}. If indeed the shift from maximal $\theta_{12}$ is due to the charged lepton diagonalization, this could offer a possible track to explain the empirical Raidal relation $\theta_{12}+\theta_C=\pi/4$ \cite{rai}(with present data $\theta_{12}+\theta_C=(47.0+1.7-1.6)^0$). While it would not be difficult in this case to arrange that the shift from maximal is of the order of $\theta_C$, it is not clear how to guarantee that it is precisely equal to $\theta_C$ \cite{smi}. Besides the effect of the charged lepton diagonalization, in a see-saw context, one can assume a strong additional breaking of $L_e-L_{\mu}-L_{\tau}$ from soft terms in the $M_{RR}$ Majorana mass matrix \cite{fra}. Since $\nu_R$'s are gauge singlets and thus essentially uncoupled, a large breaking in $M_{RR}$ does not feedback in other sectors of the lagrangian. In this way one can obtain realistic values for $\theta_{12}$ and for all other masses and mixings, in particular also with a small $\theta_{13}$.
\item[d)] Normal hierarchy. Particularly interesting are models with 23 determinant suppressed by see-saw \cite{rev}: in the 23 sector one needs relatively large mass splittings to fit the small value of $r$ but nearly maximal mixing. This can be obtained if the 23 sub-determinant is suppressed by some dynamical trick. Typical examples are lopsided models with large off diagonal term in the Dirac matrices of charged leptons and/or neutrinos (in minimal SU(5) the d-quark and charged lepton mass matrices are one the transposed of the other, so that large left-handed mixings for charged leptons correspond to large unobservable right-handed mixings for d-quarks). Another class of typical examples is the dominance in the see-saw formula of a small eigenvalue in $M_{RR}$, the right-handed Majorana neutrino mass matrix. When the 23 determinant suppression is implemented in a 33 context, normally $\theta_{13}$ is not protected from contributions that vanish with the 23 determinant, hence with $r$.
\end{itemize}

The fact that some neutrino mixing angles are large and even
nearly maximal, while surprising at the start, was eventually found to be well compatible with a unified picture of quark
and lepton masses within GUTs. The symmetry group at
$M_{GUT}$ could be either (SUSY) SU(5) or SO(10)  or a larger group. For example, normal models based on anarchy, semianarchy, inverted hierarchy or normal hierarchy can all be naturally
implemented  by simple assignments of U(1)$_{\rm F}$ horizontal charges in a semiquantitative unified
description of all quark and lepton masses in SUSY SU(5)$\times$ U(1)$_{\rm F}$. Actually, in this context, if one adopts
a statistical criterium, hierarchical models appear to be preferred over anarchy and among them normal hierarchy with see-saw appears the most likely \cite{afm}.

In conclusion we expect that experiment will eventually find that  $\theta_{13}$ is not too small and that  $\theta_{23}$ is sizably not maximal. But if, on the contrary, either $\theta_{13}$ very small or  $\theta_{23}$ very close to maximal will emerge from experiment or both, then theory will need to cope with this fact. Normal models have been extensively discussed in the literature \cite{rev}, so we concentrate here on examples of exceptional models.

\section{Tri-bimaximal Mixing}

Here we want to discuss  some particular exceptional models where both $\theta_{13}$ and $\theta_{23}- \pi/4$ exactly vanish
(more precisely,
they vanish in a suitable limit, with correction terms that can be made negligibly small) and, in addition, $s_{12}\sim1/\sqrt{3}$, a value which is in very good agreement with present data. This is the so-called tri-bimaximal or Harrison-Perkins-Scott mixing pattern  (HPS) 
\cite{hps}, with the entries in the second column all equal to $1/\sqrt{3}$ in absolute value. Here we adopt the following phase convention:
\begin{equation}
U_{HPS}= \left(\matrix{
\dd\sqrt{\frac{2}{3}}&\dd\frac{1}{\sqrt 3}&0\cr
-\dd\frac{1}{\sqrt 6}&\dd\frac{1}{\sqrt 3}&-\dd\frac{1}{\sqrt 2}\cr
-\dd\frac{1}{\sqrt 6}&\dd\frac{1}{\sqrt 3}&\dd\frac{1}{\sqrt 2}}\right)~~~~~. 
\label{2}
\end{equation}
In the HPS scheme $\tan^2{\theta_{12}}= 0.5$, to be compared with the latest experimental
determination in Table 1: $\tan^2{\theta_{12}}= 0.45^{+0.07}_{-0.04}$ (at $1\sigma$). The challenge is to find natural and appealing schemes that lead to this matrix with good accuracy. Clearly, in a natural realization of this model, a very constraining and predictive dynamics must be underlying.  It is interesting to explore particular structures giving rise to this very special set of models in a natural way. In this case we have a maximum of "order" implying special values for all mixing angles. 
Interesting ideas on how to obtain the HPS mixing matrix have been discussed in refs. \cite{hps}, \cite{continuous}, \cite{others}. 
Some attractive models 
are based on the discrete symmetry A4, which appears as particularly suitable for the purpose, and were presented 
in ref. \cite{ma1},\cite{ma1.5},\cite{ma2}, \cite{us1},\cite{us2}. 

The HPS mixing matrix suggests that mixing angles are independent of mass ratios (while for quark mixings relations like $\lambda_C^2\sim m_d/m_s$ are typical). In fact in the basis where charged lepton masses are 
diagonal, the effective neutrino mass matrix in the HPS case is given by $m_{\nu}=U_{HPS}\rm{diag}(m_1,m_2,m_3)U_{HPS}^T$:
\begin{equation}
m_{\nu}=  \left[\frac{m_3}{2}M_3+\frac{m_2}{3}M_2+\frac{m_1}{6}M_1\right]~~~~~. 
\label{1k}
\end{equation}
where:
\begin{equation}
M_3=\left(\matrix{
0&0&0\cr
0&1&-1\cr
0&-1&1}\right),
\label{2k}
\end{equation}
\begin{equation}
M_2=\left(\matrix{
1&1&1\cr
1&1&1\cr
1&1&1}\right),
\label{3k}
\end{equation}
\begin{equation}
M_1=\left(\matrix{
4&-2&-2\cr
-2&1&1\cr
-2&1&1}\right).
\label{4k}
\end{equation}
The eigenvalues of $m_{\nu}$ are $m_1$, $m_2$, $m_3$ with eigenvectors $(-2,1,1)/\sqrt{6}$, $(1,1,1)/\sqrt{3}$ and $(0,1,-1)/\sqrt{2}$, respectively. In general, disregarding possible Majorana phases, there are six parameters in a real symmetric matrix like $m_{\nu}$: here only three are left after the values of the three mixing angles have been fixed \`a la HPS. For a hierarchical spectrum $m_3>>m_2>>m_1$, $m_3^2 \sim \Delta m^2_{atm}$, $m_2^2/m_3^2 \sim \Delta m^2_{sol}/\Delta m^2_{atm}$ and $m_1$ could be negligible. But also degenerate masses and inverse hierarchy can be reproduced: for example, by taking $m_3= - m_2=m_1$  we have a degenerate model, while for $m_1= - m_2$ and $m_3=0$ an inverse hierarchy case is realized (stability under renormalization group running strongly prefers opposite signs for the first and the second eigenvalue which are related to solar oscillations and have the smallest mass squared splitting). From the general expression of the eigenvectors one immediately sees that this mass matrix, independent of the values of $m_i$, leads to the HPS mixing matrix. 

It is interesting to recall that the most general mass matrix, in the basis where charged leptons are diagonal, that corresponds to $\theta_{13}=0$ and $\theta_{23}$ maximal is of the form \cite{gri}:
\begin{equation}
m=\left(\matrix{
x&y&y\cr
y&z&w\cr
y&w&z}\right),
\label{gl}
\end{equation}
Note that this matrix is symmetric under 2-3 or $\mu - \tau$ exchange. It is however not easy to make a model where the $\mu - \tau$ applies to the whole lepton sector \cite{mutau}. Imposing the symmetry on $l^Tm_\nu l$ does not work because the Dirac mass term $ l^c m_Dl$ then produces a charged lepton mixing that completely spoils $\theta_{23}$ maximal. For example, in the model \cite{moha5}, the $\mu - \tau$ symmetry is badly broken in the charged lepton mass sector and, as a result, for parameter choices that fit the masses, $\theta_{23}$ is not necessarily close to maximal and $\theta_{13}$ is not too small: finally the model looks like a "normal" model! Similarly a symmetry $\nu_{\mu R} - \nu_{\tau R}$ in the RH neutrino sector does not lead to a $\mu - \tau$ symmetric neutrino mass after see-saw. A more elaborate broken symmetry is needed, like a set of discrete broken symmetries that make the charged lepton mass matrix diagonal and, at the same time, the Dirac neutrino mass matrix diagonal and $\mu - \tau$ symmetric and finally the permutational $2-3$ symmetry is in the RR Majorana mass matrix. Thus the idea of a "simple" $\mu - \tau$ symmetry ends up with leading to complicated models. 

For $\theta_{13}=0$ there is no CP violation, so that, disregarding Majorana phases, we can restrict our consideration to real parameters. There are four of them in eq.(\ref{gl}) which correspond to three mass eigenvalues and one remaining mixing angle, $\theta_{12}$. In particular, $\theta_{12}$ is given by:
\be \label{teta12}
\sin^2{2\theta_{12}}=\frac{8y^2}{(x-w-z)^2+8y^2}
\ee
In the HPS case $\theta_{12}$ is also fixed and an additional parameter, for example $x$, can be eliminated, leading to:
\begin{equation}
m=\left(\matrix{
z+w-y&y&y\cr
y&z&w\cr
y&w&z}\right),
\label{gl2}
\end{equation}
It is easy to see that the HPS mass matrix in eqs.(\ref{1k}-\ref{4k}) is indeed of the form in eq.(\ref{gl2}).

In the next sections we will present models of tri-bimaximal mixing based on the A4 group.
We first introduce A4 and its representations and then we show that this group is particularly suited to the problem.

\section{The A4 Group}

A4 is the group of the even permutations of 4 objects. it has 4!/2=12 elements. Geometrically, it can be seen as the invariance group of a tethraedron (the odd permutations, for example the exchange of two vertices, cannot be obtained by moving a rigid solid). Let us denote a generic permutation $(1,2,3,4)\rightarrow (n_1,n_2,n_3,n_4)$ simply by $(n_1n_2n_3n_4)$. $A4$ can be generated by two basic permutations $S$ and $T$ given by $S=(4321)$ and $T=(2314)$. One checks immediately that:
\be\label{pres}
S^2=T^3=(ST)^3=1
\ee
This is called a "presentation" of the group. The 12 even permutations belong to 4 equivalence classes ($h$ and $k$ belong to the same class if there is a $g$ in the group such that $ghg^{-1}=k$) and are generated from $S$ and $T$ as follows:
\bea \label{class}
&C1&: I=(1234)\\ \nonumber
&C2&: T=(2314),ST=(4132),\\ \nonumber ~~~~&~&~~~~TS=(3241),STS=(1423)\\ \nonumber
&C3&: T^2=(3124),ST^2=(4213),\\ \nonumber~~~~&~&~~~~T^2S=(2431),TST=(1342)\\ \nonumber
&C4&: S=(4321),T^2ST=(3412),TST^2=(2143)\\ \nonumber
\eea
Note that, except for the identity $I$ which always forms an equivalence class in itself, the other classes are according to the powers of $T$ (in C4 $S$ could as well be seen as $ST^3$).

In a finite group the squared dimensions of the inequivalent irreducible representations add up to $N$, the number of transformations in the group ($N=12$ in $A4$). $A4$ has four inequivalent representations: three of dimension one, $1$, $1'$ and $1"$ and one of dimension $3$. It is immediate to see that the one-dimensional unitary representations are obtained by:
\bea \label{uni}
1&S=1&T=1\\ \nonumber
1'&S=1&T=e^{\dd i 2 \pi/3}\equiv\omega\\\nonumber
1''&S=1&T=e^{\dd i 4\pi/3}\equiv\omega^2\\\nonumber
\eea
Note that $\omega=-1/2+\sqrt{3}/2$ is the cubic root of 1 and satisfies $\omega^2=\omega^*$, $1+\omega+\omega^2=0$.

\begin{table}[t]
\begin{center}
\caption{Characters of A4}
\begin{tabular}{|l|c|c|c|c|}
\hline \textbf{Class} & \textbf{$\chi^1$} & \textbf{$\chi^{1'}$} &\textbf{$\chi^{1"}$}&\textbf{$\chi^3$}\\
\hline $C_1$ & 1 & 1& 1&3\\
\hline $C_2$ & 1 &$\omega$&$\omega^2$&0\\
\hline $C_3$ & 1 & $\omega^2$&$\omega$&0\\
\hline $C_4$ & 1 & 1& 1&-1\\
\hline
\end{tabular}
\label{tcar}
\end{center}
\end{table}

The three-dimensional unitary representation, in a basis
where the element $S$ is diagonal, is built up from:
\begin{equation}\label{tre}
S=\left(\matrix{
1&0&0\cr
0&-1&0\cr
0&0&-1}\right),~
T=\left(\matrix{
0&1&0\cr
0&0&1\cr
1&0&0}
\right).
\ee

The characters of a group $\chi_g^R$ are defined, for each element $g$, as the trace of the matrix that maps the element in a given representation $R$. It is easy to see that equivalent representations have the same characters and that characters have the same value for all elements in an equivalence class. Characters satisfy $\sum_g \chi_g^R \chi_g^{S*}= N \delta^{RS}$. Also, for each element $h$, the character of $h$ in a direct product of representations is the product of the characters: $\chi_h^{R\otimes S}=\chi_h^R \chi_h^S$ and also is equal to the sum of the characters in each representation that appears in the decomposition of $R\otimes S$. 
The character table of A4 is given in Table II \cite{ma1}. 
From this Table one derives that indeed there are no more inequivalent irreducible representations other than $1$, $1'$, $1"$ and $3$. Also, the multiplication rules are clear: the product of two 3 gives $3 \times 3 = 1 + 1' + 1'' + 3 + 3$ and $1' \times 1' = 1''$, $1' \times 1'' = 1$, $1'' \times 1'' = 1'$ etc.
If $3\sim (a_1,a_2,a_3)$ is a triplet transforming by the matrices in eq.(\ref{tre}) we have that under $S$: $S(a_1,a_2,a_3)^t= (a_1,-a_2,-a_3)^t$ (here the upper index $t$ indicates transposition)  and under $T$: $T(a_1,a_2,a_3)^t= (a_2,a_3,a_1)^t$. Then, from two such triplets $3_a\sim (a_1,a_2,a_3)$, $3_b\sim (b_1,b_2,b_3)$ the irreducible representations obtained from their product are:
\begin{equation}
1=a_1b_1+a_2b_2+a_3b_3
\end{equation}
\begin{equation}
1'=a_1b_1+\omega^2 a_2b_2+\omega a_3b_3
\end{equation}
\begin{equation}
1"=a_1b_1+\omega a_2b_2+\omega^2 a_3b_3
\end{equation}
\begin{equation}
3\sim (a_2b_3, a_3b_1, a_1b_2)
\end{equation}
\begin{equation}
3\sim (a_3b_2, a_1b_3, a_2b_1)
\end{equation}
In fact, take for example the expression for $1"=a_1b_1+\omega a_2b_2+\omega^2 a_3b_3$. Under $S$ it is invariant and under $T$ it goes into $a_2b_2+\omega a_3b_3+\omega^2 a_1b_1=\omega^2[a_1b_1+\omega a_2b_2+\omega^2 a_3b_3]$ which is exactly the transformation corresponding to $1"$. 

In eq.(\ref{tre}) we have the representation 3 in a basis where $S$ is diagonal. It is interesting to go to a basis where instead it is $T$ which is diagonal. This is obtained through the unitary transformation:
\bea\label{trep}
T'&=&VTV^\dagger=\left(\matrix{
1&0&0\cr
0&\omega&0\cr
0&0&\omega^2}
\right),\\
S'&=&VSV^\dagger=\frac{1}{3} \left(\matrix{
-1&2&2\cr
2&-1&2\cr
2&2&-1}\right).
\eea
where:
\be \label {vu}
V=\frac{1}{\sqrt{3}} \left(\matrix{
1&1&1\cr
1&\omega^2&\omega\cr
1&\omega&\omega^2}\right).
\ee
The matrix $V$ is special in that it is a 3x3 unitary matrix with all entries of unit absolute value. It is interesting that this matrix was proposed long ago as a possible mixing matrix for neutrinos \cite{cab}. We shall see in the following that the matrix $V$ appears in $A4$ models as the unitary transformation that diagonalizes the charged lepton mass matrix .

An obvious representation of $A4$ is obtained by considering the 4x4 matrices that directly realize each permutation. For $S=(4321)$ and $T=(2314)$ we have: 
\bea\label{trepp}
S_4&=& \left(\matrix{
0& 0& 0& 1\cr
0& 0& 1& 0\cr
0& 1& 0& 0\cr
1& 0& 0& 0}
\right),\\
T_4&=&\left(\matrix{
0& 1& 0& 0\cr
0& 0& 1& 0\cr
1& 0& 0& 0\cr
0& 0& 0& 1}
\right)~~~.
\eea

The matrices $S_4$ and $T_4$ satisfy the relations (\ref{pres}), thus providing a representation
of A4. Since the only irreducible representations of A4 are a triplet and three singlets,
the 4x4 representation described by $S_4$ and $T_4$ is not irreducible. It decomposes
into the sum of the invariant singlet plus the triplet representation. 
This decomposition is realized by the unitary matrix
$U$ given by:
\be \label {u43}
U=\frac{1}{2}
\left(\matrix{
+1&+1&+1&+1\cr
-1&+1&+1&-1\cr
+1&-1&+1&-1\cr
+1&+1&-1&-1}
\right)~~~.
\ee
This matrix maps $S_4$ and $T_4$ into matrices that 
are block-diagonal:
\be
U S_4 U^\dagger=
\left(
\begin{array}{c|ccc}
1& &0&\\
\hline
& & &\\
0& &S&\\
& & &
\end{array}
\right)~~~,~~~~~~~~~~
U T_4 U^\dagger=
\left(
\begin{array}{c|ccc}
1& &0&\\
\hline
& & &\\
0& &T&\\
& & &
\end{array}
\right)~~~,
\ee
where $S$ and $T$ are the generators of the three-dimensional representation in eq.(\ref{tre}).

There is an interesting relation \cite{us2} between the $A_4$ model considered so far and the modular group. This relation could possibly be relevant to understand the origin of the A4 symmetry from a more fundamental layer of the theory.
The modular group $\Gamma$ is the group of linear fractional transformations acting on a complex variable $z$:
\be
z\to\frac{az+b}{cz+d}~~~,~~~~~~~ad-bc=1~~~,
\label{frac}
\ee
where $a,b,c,d$ are integers. 
There are infinite elements in $\Gamma$, but all of them can be generated by the two
transformations:
\be
s:~~~z\to -\frac{1}{z}~~~,~~~~~~~t:~~~z\to z+1~~~,
\label{st}
\ee
The transformations $s$ and $t$ in (\ref{st}) satisfy the relations
\be
s^2=(st)^3=1
\label{absdef}
\ee
and, conversely, these relations provide an abstract characterization of the modular group.
Since the relations (\ref{pres}) are a particular case of the more general constraint (\ref{absdef}),
it is clear that A4 is a very small subgroup of the modular group and that the A4 representations discussed above are also representations of the modular group.
In string theory the transformations (\ref{st})
operate in many different contexts. For instance the role of the complex 
variable $z$ can be played by a field, whose VEV can be related to a physical
quantity like a compactification radius or a coupling constant. In that case
$s$ in eq. (\ref{st}) represents a duality transformation and $t$ in eq. (\ref{st}) represent the transformation associated to an ''axionic'' symmetry.

\section{Applying A4 to Lepton Masses and Mixings}

A typical A4 model works as follows \cite{us1},  \cite{us2}. One assigns
leptons to the four inequivalent
representations of A4: left-handed lepton doublets $l$ transform
as a triplet $3$, while the right-handed charged leptons $e^c$,
$\mu^c$ and $\tau^c$ transform as $1$, $1'$ and $1''$, respectively. 
At this stage we do not introduce RH neutrinos, but later we will discuss a see-saw realization. The flavour symmetry is broken by two real triplets
$\varphi$ and $\varphi'$ and by a real singlet $\xi$. 
These flavon fields are all gauge singlets.
We also need one or two ordinary SM Higgs doublets $h_{u,d}$, which we take invariant under A4. 
The Yukawa interactions in the lepton sector read:
\bea \label{ltt}
{\cal L}_Y&=&y_e e^c (\varphi l)+y_\mu \mu^c (\varphi l)''+
y_\tau \tau^c (\varphi l)'\\ \nonumber
&+& x_a\xi (ll)+x_d (\varphi' ll)+h.c.+...
\eea
In our notation, $(3 3)$ transforms as $1$, 
$(3 3)'$ transforms as $1'$ and $(3 3)''$ transforms as $1''$.
Also, to keep our notation compact, we use a two-component notation
for the fermion fields and we set to 1 the Higgs fields
$h_{u,d}$ and the cut-off scale $\Lambda$. For instance 
$y_e e^c (\varphi l)$ stands for $y_e e^c (\varphi l) h_d/\Lambda$,
$x_a\xi (ll)$ stands for $x_a\xi (l h_u l h_u)/\Lambda^2$ and so on.
The Lagrangian  ${\cal L}_Y$ contains the lowest order operators
in an expansion in powers of $1/\Lambda$. Dots stand for higher
dimensional operators that will be discussed later. 
Some terms allowed by the flavour symmetry, such as the terms 
obtained by the exchange $\varphi'\leftrightarrow \varphi$, 
or the term $(ll)$ are missing in ${\cal L}_Y$. 
Their absence is crucial and, in each version of A4 models, is
motivated by additional symmetries. For example $(ll)$,  being of lower dimension
with respect to $ (\varphi' ll)$,  would be the dominant component, proportional to the identity, of the neutrino mass matrix. In addition to that, the presence of the singlet flavon $\xi$ plays an important role in making the VEV directions of $\varphi$ and $\varphi'$ different.

For the model to work it is essential that the fields $\varphi'$,
$\varphi$ and $\xi$ develop a VEV along the directions:
\bea
\langle \varphi' \rangle&=&(v',0,0)\nn\\ 
\langle \varphi \rangle&=&(v,v,v)\nn\\
\langle \xi \rangle&=&u~~~. 
\label{align}
\eea 
A crucial part of all serious A4 models is the dynamical generation of this alignment in a natural way. If the alignment is realized, at the leading order of the $1/\Lambda$ expansion,
the mass matrices $m_l$ and $m_\nu$ for charged leptons and 
neutrinos are given by:
\be
m_l=v_d\frac{v}{\Lambda}\left(
\begin{array}{ccc}
y_e& y_e& y_e\\
y_\mu& y_\mu \omega^2& y_\mu \omega\\
y_\tau& y_\tau \omega& y_\tau \omega^2
\end{array}
\right)~~~,
\label{mch}
\ee
\be
m_\nu=\frac{v_u^2}{\Lambda}\left(
\begin{array}{ccc}
a& 0& 0\\
0& a& d\\
0& d& a
\end{array}
\right)~~~,
\label{mnu}
\ee
where
\be
a\equiv x_a\frac{u}{\Lambda}~~~,~~~~~~~d\equiv x_d\frac{v'}{\Lambda}~~~.
\label{ad}
\ee
Charged leptons are diagonalized by the matrix
\be
l\to V l =\frac{1}{\sqrt{3}}\left(
\begin{array}{ccc}
1& 1& 1\\
1& \omega^2& \omega\\
1& \omega& \omega^2
\end{array}
\right)l~~~,
\label{change}
\ee
This matrix was already introduced in eq.(\ref{vu}) as the unitary transformation between the $S$-diagonal to the $T$-diagonal 3x3 representation of $A4$. In fact, in this model, the $S$-diagonal basis is the Lagrangian basis and the $T$ diagonal basis is that of diagonal charged leptons. The great virtue of $A4$ is to immediately produce the special unitary matrix $V$ as the diagonalizing matrix of charged leptons and also to allow a singlet made up of three triplets, $(\phi' ll) = \phi'_1l_2l_3+\phi'_2l_3l_1+\phi'_3l_1l_2$ which leads, for the alignment in eq. (\ref{align}), to the right neutrino mass matrix to obtain the HPS mixing matrix.

The charged fermion masses are given by:
\bea \label{chmasses}
m_e=\sqrt{3} y_e v_d \frac{v}{\Lambda}~~~,~~~~~~~
m_\mu&=&\sqrt{3} y_\mu v_d \frac{v}{\Lambda}~~~,~~~~~~~\nonumber \\
m_\tau&=&\sqrt{3} y_\tau v_d \frac{v}{\Lambda}~~~.
\eea
We can easily obtain in a a natural way the observed hierarchy among $m_e$, $m_\mu$ and
$m_\tau$ by introducing an additional U(1)$_F$ flavour symmetry under
which only the right-handed lepton sector is charged.
We assign F-charges $0$, $2$ and $3\div 4$ to $\tau^c$, $\mu^c$ and
$e^c$, respectively. By assuming that a flavon $\theta$, carrying
a negative unit of F, acquires a VEV 
$\langle \theta \rangle/\Lambda\equiv\lambda<1$, the Yukawa couplings
become field dependent quantities $y_{e,\mu,\tau}=y_{e,\mu,\tau}(\theta)$
and we have
\be
y_\tau\approx O(1)~~~,~~~~~~~y_\mu\approx O(\lambda^2)~~~,
~~~~~~~y_e\approx O(\lambda^{3\div 4})~~~.
\ee
In the flavour basis the neutrino mass matrix reads [notice that the change of basis induced by $V$, because of the Majorana nature of neutrinos,
will in general change (compare eq.(\ref{mnu}) with eq.(\ref{mnu0})) the relative phases of the eigenvalues of $m_\nu$]:
\be
m_\nu=\frac{v_u^2}{\Lambda}\left(
\begin{array}{ccc}
a+2 d/3& -d/3& -d/3\\
-d/3& 2d/3& a-d/3\\
-d/3& a-d/3& 2 d/3
\end{array}
\right)~~~,
\label{mnu0}
\ee
and is diagonalized by the transformation:
\be
U^T m_\nu U =\frac{v_u^2}{\Lambda}{\tt diag}(a+d,a,-a+d)~~~,
\ee
with
\be
U=\left(
\begin{array}{ccc}
\sqrt{2/3}& 1/\sqrt{3}& 0\\
-1/\sqrt{6}& 1/\sqrt{3}& -1/\sqrt{2}\\
-1/\sqrt{6}& 1/\sqrt{3}& +1/\sqrt{2}
\end{array}
\right)~~~.
\ee
The leading order predictions are $\tan^2\theta_{23}=1$, 
$\tan^2\theta_{12}=0.5$ and $\theta_{13}=0$. The neutrino masses
are $m_1=a+d$, $m_2=a$ and $m_3=-a+d$, in units of $v_u^2/\Lambda$.
We can express $|a|$, $|d|$ in terms of 
$r\equiv \Delta m^2_{sol}/\Delta m^2_{atm}
\equiv (|m_2|^2-|m_1|^2)/|m_3|^2-|m_1|^2)$,
$\Delta m^2_{atm}\equiv|m_3|^2-|m_1|^2$ 
and $\cos\Delta$, $\Delta$ being the phase difference between
the complex numbers $a$ and $d$:
\bea
\sqrt{2}|a|\frac{v_u^2}{\Lambda}&=&
\frac{-\sqrt{\Delta m^2_{atm}}}{2 \cos\Delta\sqrt{1-2r}}\nn\\
\sqrt{2}|d|\frac{v_u^2}{\Lambda}&=&
\sqrt{1-2r}\sqrt{\Delta m^2_{atm}}~~~.
\label{tuning}
\eea
To satisfy these relations a moderate tuning is needed in this model.
Due to the absence of $(ll)$ in eq. (\ref{ltt}) which we will motivate in the next section, $a$ and $d$ are of the same order in $1/\Lambda$, 
see eq. (\ref{ad}). Therefore we expect that $|a|$ and $|d|$ 
are close to each other and, to satisfy eqs. (\ref{tuning}),
$\cos\Delta$ should be negative and of order one. We obtain:
\bea
|m_1|^2&=&\left[-r+\frac{1}{8\cos^2\Delta(1-2r)}\right]
\Delta m^2_{atm}\nn\\
|m_2|^2&=&\frac{1}{8\cos^2\Delta(1-2r)}
\Delta m^2_{atm}\nn\\
|m_3|^2&=&\left[1-r+\frac{1}{8\cos^2\Delta(1-2r)}\right]\Delta m^2_{atm}
\label{lospe}
\eea
If $\cos\Delta=-1$, we have a neutrino spectrum close to hierarchical:
\be
|m_3|\approx 0.053~~{\rm eV}~~~,~~~~~~~
|m_1|\approx |m_2|\approx 0.017~~{\rm eV}~~~.
\ee 
In this case the sum of neutrino masses is about $0.087$ eV.
If $\cos\Delta$ is accidentally small, the neutrino spectrum becomes
degenerate. The value of $|m_{ee}|$, the parameter characterizing the 
violation of total lepton number in neutrinoless double beta decay,
is given by:
\be
|m_{ee}|^2=\left[-\frac{1+4 r}{9}+\frac{1}{8\cos^2\Delta(1-2r)}\right]
\Delta m^2_{atm}~~~.
\ee
For $\cos\Delta=-1$ we get $|m_{ee}|\approx 0.005$ eV, at the upper edge of
the range allowed for normal hierarchy, but unfortunately too small
to be detected in a near future.
Independently from the value of the unknown phase $\Delta$
we get the relation:
\be
|m_3|^2=|m_{ee}|^2+\frac{10}{9}\Delta m^2_{atm}\left(1-\frac{r}{2}\right)~~~,
\ee
which is a prediction of this model.

\section{A4 model with an extra dimension}
One of the problems we should solve in the quest for
the correct alignment is that of keeping neutrino and charged
lepton sectors separate, allowing $\varphi$ and $\varphi'$ to take different VEVs and also forbidding the exchange of one with the other in interaction terms. One possibility is that this separation is achieved
by means of an extra spatial dimension. The space-time is assumed
to be five-dimensional, the product of the four-dimensional
Minkowski space-time times an interval going from $y=0$ to $y=L$. 
At $y=0$ and $y=L$ the space-time has two four-dimensional boundaries,
called "branes". The idea is that matter SU(2) singlets
such as $e^c,\mu^c,\tau^c$ are localized at $y=0$, while SU(2) doublets,
such as $l$ are localized at $y=L$ (see Fig.1). Neutrino masses
arise from local operators at $y=L$. Charged lepton
masses are produced by non-local effects involving both branes.
The simplest possibility is to introduce a bulk fermion,
depending on all space-time coordinates, that interacts with
$e^c,\mu^c,\tau^c$ at $y=0$ and with $l$ at $y=L$. The exchange of
such a fermion can provide the desired non-local coupling between
right-handed and left-handed ordinary fermions. Finally,
assuming that $\varphi$ and $(\varphi',\xi)$ are localized
respectively at $y=0$ and $y=L$, one obtains a natural separation
between the two sectors.

\begin{figure}[]
\centering
$$\hspace{-4mm}
\includegraphics[width=6.0 cm]{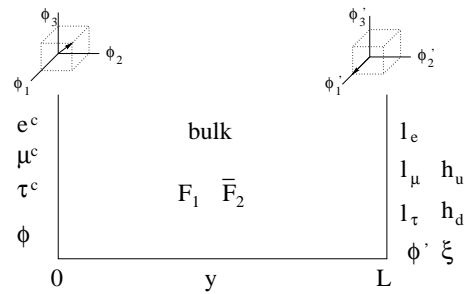}
$$
\caption[]
{Fifth dimension and localization of scalar and fermion fields.
The symmetry breaking sector includes the A4 triplets $\varphi$
and $\varphi'$, localized at the opposite ends of the interval.
Their VEVs are dynamically aligned along the directions shown
at the top of the figure.}
\end{figure}

Such a separation also greatly simplifies the vacuum alignment
problem. One can determine the minima of two
scalar potentials $V_0$ and $V_L$, depending only, respectively,
on $\varphi$ and $(\varphi',\xi)$. Indeed, it is shown that there are whole regions of the parameter space where 
$V_0(\varphi)$ and $V_L(\varphi',\xi)$
have the minima given in eq. (\ref{align}). Notice that in the present setup
dealing with a discrete symmetry such as A4 provides a 
great advantage as far as the alignment problem 
is concerned. A continuous flavour symmetry such as, for instance,
SO(3) would need some extra structure to achieve the desired
alignment. Indeed the potential energy 
$\int d^4x [V_0(\varphi)+V_L(\varphi',\xi)]$ would be
invariant under a much bigger symmetry, SO(3)$_0\times$ SO(3)$_L$,
with the SO(3)$_0$ acting on $\varphi$ and leaving $(\varphi',\xi)$
invariant and vice-versa for SO(3)$_L$.
This symmetry would remove any alignment between the VEVs
of $\varphi$ and those of $(\varphi',\xi)$.
If, for instance, (\ref{align}) is minimum of the potential energy,
then any other configuration obtained by acting on (\ref{align})
with SO(3)$_0\times$ SO(3)$_L$ would also be a minimum
and the relative orientation between the two sets of VEVs would be
completely undetermined.
A discrete symmetry such as A4 has not this problem, because applying separate A4 transformation on the minimum solutions on each brane a finite number of degenerate vacua is obtained which can be shown to correspond to the same physics apart from redefinitions of fields and parameters.

The Lagrangian in 5 dimensions includes a bulk fermion field $F(x,y)=(F_1,\overline{F_2})$,
singlet under SU(2) with hypercharge $Y=-1$ and transforming
as a triplet of A4. One also imposes a discrete $Z_4$
symmetry under which $(f^c,l,F,\varphi,
\varphi',\xi)$ transform into 
$(-i f^c,i l,i F,\varphi,-\varphi',-\xi)$. The complete action is
\begin{widetext}
\bea
S&=&\int d^4x dy \left\{
\left[
iF_1 \sigma^\mu \partial_\mu \ov{F}_1
+iF_2 \sigma^\mu \partial_\mu \ov{F}_2
+\frac{1}{2}(F_2\partial_y F_1-\partial_y F_2 F_1+h.c.)
\right]\right.\nn\\
&-&M(F_1F_2+\ov{F}_1\ov{F}_2)\nn\\
&+&V_0(\varphi)\delta(y)+V_L(\varphi',\xi)\delta(y-L)\nn\\
&+&\left[Y_e e^c (\varphi F_1)+Y_\mu \mu^c (\varphi F_1)''+
Y_\tau \tau^c (\varphi F_1)'+h.c.\right]\delta(y)\nn\\
&+&\left.\left[\frac{x_a}{\Lambda^2}\xi (ll)h_u h_u+
\frac{x_d}{\Lambda^2} (\varphi' ll)h_u h_u
+Y_L(F_2 l)h_d +h.c.\right]\delta(y-L)\right\}+...~~~,
\label{action5}
\eea
\end{widetext}
where the constants $Y$ have mass dimension -1/2.
The first two lines represent the five-dimensional kinetic and mass terms
of the bulk field $F$. The third line is the scalar potential 
and the remaining terms are the lowest order invariant operators
localized at the two branes. Dots stand for the kinetic terms
of $f^c,l,\varphi,\varphi',\xi$ and for higher-dimensional
operators.

The potential energy is given, at lowest order by:
\be
U=\int d^4x\left[V_0(\varphi)+V_L(\varphi',\xi)\right]~~~,
\label{penergy}
\ee
and, under the conditions discussed above, is minimized by
eqs. (\ref{align}) \cite{us1}.
It is clear that at lowest order $\varphi$ and $(\varphi',\xi)$ are strictly 
separated. 

We now discuss the effects of the tree-level exchange of $F$.
To this purpose we consider the equations of motion for $(F_1,F_2)$:
\bea
i \sigma^\mu\partial_\mu\ov{F}_2+\partial_y F_1-M F_1&=&0\nn\\
i \sigma^\mu\partial_\mu\ov{F}_1-\partial_y F_2-M F_2&=&0
\eea
If $M$ is large and positive, we can prove that all the modes
contained in $(F_1,F_2)$ become heavy, at a scale greater than or comparable
to $1/L$, which we assume to be much higher than the electroweak scale.
If we are only interested in energies much lower than $1/L$,
we can solve the equations of motion in the static approximation,
by neglecting the four-dimensional kinetic term:
\bea
F_1(y)&=&F_1(L) e^{M(y-L)}\nn\\
F_2(y)&=&F_2(0) e^{-M y}~~~.
\label{sol}
\eea
These equations must be supplemented with appropriate boundary conditions,
which can be identified by varying the action $S$ with respect the fields
$(F_1,F_2)$. As a final result, as shown in detail in ref. \cite{us1}, in lowest order approximation the Lagrangian
${\cal L}_Y$ of eq. (\ref{ltt}) is reproduced and the general discussion
applies.

We also recall that, to account for the observed hierarchy
of the charged lepton masses, one has to include an additional
U(1) flavour symmetry. Therefore, in the present picture, the
quantities $Y_{e,\mu,\tau}$ stand for:
\be
Y_e=\tilde{Y}_e \left(\frac{\theta}{\Lambda}\right)^4~~~~~~~,~~~
Y_\mu=\tilde{Y}_\mu \left(\frac{\theta}{\Lambda}\right)^2~~~~~~~,~~~
Y_\tau=\tilde{Y}_\tau~~~,
\ee
where $\tilde{Y}_{e,\mu,\tau}$ are field-independent constants
having similar values. After spontaneous breaking of $U(1)$,
the Yukawa couplings $y_f$ possess the desired hierarchy.

\section{A4 model with SUSY in 4 Dimensions}

We now discuss an alternative supersymmetric solution to the vacuum alignment problem \cite{us2}.
In a SUSY context, the right-hand side of eq. (\ref{ltt})
should be interpreted as the superpotential $w_l$ of the theory,
in the lepton sector:
\bea \label{wlplus}
w_l&=&y_e e^c (\varphi l)+y_\mu \mu^c (\varphi l)"+
y_\tau \tau^c (\varphi l)'+\\ \nonumber&+& (x_a\xi+\tilde{x}_a\tilde{\xi}) (ll)+x_b (\varphi' ll)+h.c.+...
\eea
where dots stand for higher dimensional operators  and where we have also added an additional 
A4-invariant singlet
$\tilde{\xi}$. Such a singlet does not modify the structure of the mass matrices
discussed previously, but plays an important role in the vacuum 
alignment mechanism.
A key observation is that the superpotential $w_l$ 
is invariant not only with respect to the gauge symmetry 
SU(2)$\times$ U(1) and the flavour symmetry U(1)$_F\times A_4$,
but also under a discrete $Z_3$ symmetry and a continuous U(1)$_R$ 
symmetry under which the fields 
transform as shown in the following table.
\\[0.2cm]
\begin{center}
\begin{tabular}{|c||c|c|c|c||c|c|c|c|c||c|c|c|}
\hline
{\tt Field}& l & $e^c$ & $\mu^c$ & $\tau^c$ & $h_{u,d}$ & 
$\varphi$ & $\varphi'$ & $\xi$ & $\tilde{\xi}$ & $\varphi_0$ & $\varphi_0'$ & $\xi_0$\\
\hline
A4 & $3$ & $1$ & $1'$ & $1''$ & $1$ & 
$3$ & $3$ & $1$ & $1$ & $3$ & $3$ & $1$\\
\hline
$Z_3$ & $\omega$ & $\omega^2$ & $\omega^2$ & $\omega^2$ & $1$ &
$1$ & $\omega$ & $\omega$ & $\omega$ & $1$ & $\omega$ & $\omega$\\
\hline
$U(1)_R$ & $1$ & $1$ & $1$ & $1$ & $0$ & 
$0$ & $0$ & $0$ & $0$ & $2$ & $2$ & $2$\\
\hline
\end{tabular}
\end{center}
\vspace{0.2cm}
We see that the $Z_3$ symmetry explains the absence of the term $(ll)$
in $w_l$: such a term transforms as $\omega^2$ under $Z_3$ and
need to be compensated by the field $\xi$ in our construction.
At the same time $Z_3$ does not allow the interchange between
$\varphi$ and $\varphi'$, which transform differently under $Z_3$. 
The singlets $\xi$ and $\tilde{\xi}$ have the same transformation properties
under all symmetries and, as we shall see, in a finite range of parameters,
the VEV of $\tilde{\xi}$ vanishes and does not contribute to neutrino masses. 
Charged leptons and neutrinos acquire
masses from two independent sets of fields. 
If the two sets of fields develop VEVs according to the 
alignment described in eq. (\ref{align}), then the desired
mass matrices follow.

Finally, there is a continuous $U(1)_R$
symmetry that contains the usual $R$-parity as a subgroup.
Suitably extended to the quark sector, this symmetry forbids
the unwanted dimension two and three terms in the superpotential
that violate baryon and lepton number at the renormalizable level. 
The $U(1)_R$ symmetry allows us to classify
fields into three sectors. There are ``matter fields'' such as the 
leptons $l$, $e^c$, $\mu^c$ and $\tau^c$, which occur in the 
superpotential through bilinear combinations. There is a 
``symmetry breaking sector'' including the higgs doublets
$h_{u,d}$ and the flavons $\varphi$, $\varphi'$, $(\xi,\tilde{\xi})$.
Finally, there are ``driving fields'' such as $\varphi_0$, $\varphi_0'$
and $\xi_0$ that allows to build a 
non-trivial scalar potential in the symmetry breaking sector. 
Since driving fields have R-charge equal to two, the superpotential
is linear in these fields.

The full superpotential of the model is
\be
w=w_l+w_d
\ee 
where, at leading order in a $1/\Lambda$ expansion, $w_l$ is given
by  eq. (\ref{wlplus}) and the ``driving'' term 
$w_d$ reads:
\bea
w_d&=&M (\varphi_0 \varphi)+ g (\varphi_0 \varphi\varphi)\nn\\
&+&g_1 (\varphi_0' \varphi'\varphi')+
g_2 \tilde{\xi} (\varphi_0' \varphi')+
g_3 \xi_0 (\varphi'\varphi')\nn\\
&+&
g_4 \xi_0 \xi^2+
g_5 \xi_0 \xi \tilde{\xi}+
g_6 \xi_0 \tilde{\xi}^2~~~.
\label{wd}
\eea
At this level there is no fundamental distinction between the singlets
$\xi$ and $\tilde{\xi}$. Thus we are free to define $\tilde{\xi}$ as the combination
that couples to $(\varphi_0' \varphi')$ in the superpotential $w_d$.
We notice that at the leading order there are no terms involving
the Higgs fields $h_{u,d}$. We assume that the electroweak symmetry
is broken by some mechanism, such as radiative effects when SUSY is broken. It is interesting that at the leading order
the electroweak scale does not mix with the potentially large scales
$u$, $v$ and $v'$. The scalar potential is given by:
\be
V=\sum_i\left\vert\frac{\partial w}{\partial \phi_i}\right\vert^2
+m_i^2 \vert \phi_i\vert^2+...
\ee
where $\phi_i$ denote collectively all the scalar fields of the 
theory, $m_i^2$ are soft masses and dots stand for D-terms for the 
fields charged under the gauge group and possible additional
soft breaking terms. Since $m_i$ are expected to be much smaller
than the mass scales involved in $w_d$, it makes sense to
minimize $V$ in the supersymmetric limit and to account for soft 
breaking effects subsequently. A detailed minimization analysis, presented in ref.\cite{us2}, shows the the desired alignment solution is indeed realized. 

\section{Corrections to the Lowest Approximation}

The results of the previous sections hold to first approximation.
Higher-dimensional operators, suppressed by additional powers of 
the cut-off $\Lambda$, can be added to the leading terms in the lagrangian.
These
corrections have been classified and discussed in detail in refs.\cite{us1}, \cite{us2}. They are completely under control in our models and can be made negligibly small without any fine-tuning: one only needs to assume that the VEV's are sufficiently smaller than the cutoff $\Lambda$.
Higher-order operators
contribute corrections to the charged lepton masses, to the neutrino mass matrix and to the vacuum alignment. These corrections, suppressed by powers of VEVs/$\Lambda$, with different exponents in different versions of A4 models, affect all the relevant observable with terms of the same order: $s_{13}$, $s_{12}$, $s_{23}$, $r$.  If we require that the subleading
terms do not spoil the leading order picture, these deviations should not be larger than
about 0.05. This can be inferred by the agreement of  the HPS value of $\tan^2\theta_{12}$ with the experimental value, from the present bound on $\theta_{13}$ or from requiring that the corrections do not exceed the measured value of $r$. In the SUSY model, where the largest corrections are linear in VEVs/$\Lambda$ \cite{us2}, this implies the bound
\be
\frac{v_S}{\Lambda}\approx \frac{v_T}{\Lambda}\approx \frac{u}{\Lambda}<0.05
\label{range2}
\ee
which does not look unreasonable, for example if VEVs$\sim M_{GUT}$ and $\Lambda\sim M_{Planck}$.

\section{See-saw Realization}
We can easily modify the previous model to implement the see-saw mechanism.
We introduce conjugate right-handed neutrino fields $\nu^c$ transforming as a triplet of A4
and we modify the transformation law of the other fields according to the following table:
\\[0.2cm]
\begin{center}
\begin{tabular}{|c||c||c|c|c||c|c|}
\hline
{\tt Field}& $\nu^c$ & $\varphi'$ & $\xi$ & $\tilde{\xi}$ & $\varphi_0'$ & $\xi_0$\\
\hline
A4 & $3$ & $3$ & $1$ & $1$ & $3$ & $1$\\ 
\hline
$Z_3$ & $\omega^2$ & $\omega^2$ & $\omega^2$ & $\omega^2$ &  $\omega^2$ & $\omega^2$\\
\hline
$U(1)_R$ & $1$ & $0$ & $0$ & $0$ & $2$ & $2$\\
\hline
\end{tabular}
\end{center}
\vspace{0.2cm}
The superpotential becomes
\be
w=w_l+w_d
\ee 
where the `driving' part is unchanged, whereas $w_l$ is now given by:
\bea \label{wlss}
w_l&=&y_e e^c (\varphi l)+y_\mu \mu^c (\varphi l)"+
y_\tau \tau^c (\varphi l)'+\\ \nonumber&+& y (\nu^c l)+
(x_A\xi+\tilde{x}_A\tilde{\xi}) (\nu^c\nu^c)\\ \nonumber&+&x_B (\varphi' \nu^c\nu^c)+h.c.+...
\eea
dots denoting higher-order contributions. The vacuum alignment proceeds exactly as
discussed in section 8 and also the charged lepton sector is unaffected by the modifications.
In the neutrino sector, after electroweak and A4 symmetry breaking we have Dirac
and Majorana masses:
\be
m^D_\nu=y v_u {\bf 1},~~
M=\left(
\begin{array}{ccc}
A& 0& 0\\
0& A& B\\
0&B& A
\end{array}
\right) u ~~~,
\ee
where ${\bf 1}$ is the unit 3$\times$3 matrix and 
\be
A\equiv 2 x_A ~~~,~~~~~~~B\equiv 2 x_B \frac{v_S}{u}~~~.
\label{add}
\ee
The mass matrix for light neutrinos is $m_\nu=(m^D_\nu)^T M^{-1} m^D_\nu$ with eigenvalues
\bea
m_1&=&\frac{y^2}{A+B}\frac{v_u^2}{u}~~~,~~~~~~~
m_2=\frac{y^2}{A}\frac{v_u^2}{u}\\ \nonumber
m_3&=&\frac{y^2}{-A+B}\frac{v_u^2}{u}~~~.
\eea
The mixing matrix is the HPS one, eq. (\ref{2}).
In the presence of a see-saw mechanism both normal and inverted hierarchies 
in the neutrino mass spectrum can be realized. If we call $\Phi$ the relative phase
between the complex number $A$ and $B$, then $\cos\Phi>-|B|/2|A|$
is required to have $|m_2|>|m_1|$. In the interval
$-|B|/2|A|<\cos\Phi\le 0$, the spectrum is of inverted hierarchy type,
whereas in $|B|/2|A|\le \cos\Phi\le 1$ the neutrino hierachy is of normal type.
It is interesting that this model is an example of model with inverse hierarchy, realistic $\theta_{12}$ and $\theta_{23}$ and, at least in a first approximation, $\theta_{13}=0$. 
The quantity $|B|/2|A|$ cannot be too large, otherwise the ratio $r$
cannot be reproduced. When $|B|\ll|A|$ the spectrum is quasi degenerate.
When $|B|\approx |A|$ we obtain the strongest hierarchy.
For instance, if $B=-2A+z$ ($|z|\ll |A|,|B|$), we find the following
spectrum:
\bea
|m_1|^2&\approx& \Delta m_{atm}^2(\frac{9}{8}+\frac{1}{12} r),\\ \nonumber
|m_2|^2&\approx& \Delta m_{atm}^2(\frac{9}{8}+\frac{13}{12} r),\\ \nonumber
|m_3|^2&\approx& \Delta m_{atm}^2(\frac{1}{8}+\frac{1}{12} r).
\eea
When $B=A+z$ ($|z|\ll |A|,|B|$), we obtain:
\bea
|m_1|^2&\approx& \Delta m_{atm}^2(\frac{1}{3} r),\\ \nonumber
|m_2|^2&\approx& \Delta m_{atm}^2(\frac{4}{3} r),\\ \nonumber
|m_3|^2&\approx&\Delta m_{atm}^2(1-\frac{1}{3} r).
\eea
These results are affected by higher-order corrections induced
by non renormalizable operators with similar results as in the version with no see-saw. In conclusion, the symmetry structure of the model is fully compatible
with the see-saw mechanism.

%
\section{Quarks}
There are several possibilities to include quarks. At first sight the most appealing
one is to adopt for quarks the same classification scheme under A4 that we have 
used for leptons. Thus we tentatively assume that left-handed quark doublets $q$ transform
as a triplet $3$, while the right-handed quarks $(u^c,d^c)$,
$(c^c,s^c)$ and $(t^c,b^c)$ transform as $1$, $1'$ and $1"$, respectively. We can similarly
extend to quarks the transformations of $Z_3$ and U(1)$_R$ given for leptons in the table
of section 8. The superpotential for quarks reads:
\bea \label{qtt}
{w}_{q}&=&
y_d d^c (\varphi q)+y_s s^c (\varphi q)"+y_b b^c (\varphi q)'\\ \nonumber
&+&
y_u u^c (\varphi q)+y_c c^c (\varphi q)"+y_t t^c (\varphi q)'
+h.c.+...
\eea
It is interesting to note that such an extrapolation to quarks leads to a diagonal CKM 
mixing matrix in first approximation \cite{ma1,ma1.5,us2, ma3}. In fact, starting from eq. (\ref{qtt}) and proceeding as 
described in detail for the lepton sector, we  see that the up quark and down quark mass matrices 
are separately diagonal with mass eigenvalues which are left unspecified by A4 and with a hierarchy 
that could be accomodated by a suitable U(1)$_F$ set of charge assignments for quarks. Thus 
the $V_{CKM}$ matrix is the identity in leading order, providing a good
first order approximation.

The problems come when we discuss non-leading corrections. As seen in section 9, 
first-order corrections to the lepton sector should be typically below 0.05,
approximately the square of the Cabibbo angle. 
Also, by inspecting these corrections more closely, we see that, up to very small terms
of order $y_{u(d)}^2/y_{t(b)}^2$ and $y_{c(s)}^2/y_{t(b)}^2$, all corrections
are the same in the up and down sectors and therefore they almost
exactly cancel in the mixing matrix $V_{CKM}$. We conclude that, if one insists in 
adopting for quarks the same flavour properties as for leptons, than
new sources of A4 breaking are needed in order to produce an acceptable $V_{CKM}$.   

The A4 classification for quarks and leptons discussed in this section, which leads to an appealing first approximation with $V_{CKM}\sim 1$ for quark mixing and to $U_{HPS}$ for neutrino mixings, is not compatible with A4 commuting with SU(5) or SO(10). In fact for this to be true all particles in a representation of SU(5) should have the same A4 classification. But, for example,  both the $Q=(u,d)_L$ LH quark doublet and the RH charged leptons $l^c$ belong to the 10 of SU(5), yet they have different A4 transformation properties.
In a recent paper \cite{maGUT} the possibility of classifying all fermion multiplets as triplets was advanced. But the crucial issues of the correct alignment and of reproducing in a natural way the observed hierarchy of, for example, the charged leptons were not addressed and are difficult to realize in this case. 

\section{Conclusion}

In the last decade we have learnt a lot about neutrino masses and mixings.  A list of important conclusions have been reached. Neutrinos are not all massless but their masses are very small. Probably masses are small because neutrinos are Majorana particles
with masses inversely proportional to the large scale M of lepton number violation. It is quite remarkable that M is empirically close to $10^{14-15} GeV$ not far from $M_{GUT}$, so that
neutrino masses fit well in the SUSY GUT picture. Also out of equilibrium decays with CP and L violation of heavy RH neutrinos can produce a B-L asymmetry, then converted near the weak scale by instantons into an amount of B asymmetry compatible with observations (baryogenesis via leptogenesis) \cite{bpy}.  It has been established that neutrinos are not a significant component of dark matter in the Universe. We have also understood there there is no contradiction between large neutrino mixings and small quark mixings, even in the context of GUTÕs.  

This is a very impressive list of achievements. Coming to a detailed analysis of neutrino masses and mixings a very long collection of models have been formulated over the years. 
With a continuous improvement of the data and a progressive narrowing of the values of the mixing angles most of the models have been discarded by experiment. Still the missing elements in the picture like, for example, the scale of the average neutrino $m^2$, the pattern of the spectrum (degenerate or inverse or normal hierarchy) and the value of $\theta_{13}$ have left many different viable alternatives for models. It certainly is a reason of satisfaction that so much has been learnt recently from experiments on neutrino mixings. By now, besides the detailed knowledge of the entries of the $V_{CKM}$ matrix we also have a reasonable determination of the neutrino mixing matrix. It is remarkable that neutrino and  quark mixings have such a different qualitative pattern. One could have imagined that neutrinos would bring a decisive boost towards the formulation of a comprehensive understanding of fermion masses and mixings. In reality it is frustrating that no real illumination was sparked on the problem of flavour. We can reproduce in many different ways the observations but we have not yet been able to single out a unique and convincing baseline for the understanding of fermion masses and mixings. In spite of many interesting ideas and the formulation of many elegant models, some of them presented in these lectures, the mysteries of the flavour structure of the three generations of fermions have not been much unveiled.

\begin{acknowledgments}

It is a very pleasant duty for me to most warmly thank the Organizers of the School for their kind invitation and for the great hospitality offered to me in Iran.
This work has been partly supported by by the European Commission
under contract MRTN-CT-2004-503369.

\end{acknowledgments}


\end{document}